\documentclass[lettersize,journal]{IEEEtran}
\usepackage{amsmath,amsfonts}
\usepackage{algorithmic}
\usepackage{algorithm}
\usepackage{array}
\usepackage[caption=false,font=normalsize,labelfont=sf,textfont=sf]{subfig}
\usepackage{textcomp}
\usepackage{stfloats}
\usepackage{url}
\usepackage{verbatim}
\usepackage{graphicx}
\usepackage{cite}
\begin{document}

\title{Adoption and Impact of ChatGPT in Computer Science Education: A Case Study on a Database Administration Course}

\author{Daniel López-Fernández, and Ricardo Vergaz
\thanks{Daniel López-Fernández and Ricardo Vergaz are, respectively, with Universidad Politécnica de Madrid and Universidad Carlos III de Madrid}
\thanks{Manuscript received April 19, 2021; revised August 16, 2021.}}

\markboth{Arxiv}%
{Shell \MakeLowercase{\textit{et al.}}: A Sample Article Using IEEEtran.cls for IEEE Journals}

\IEEEpubid{0000--0000/00\$00.00~\copyright~2021 IEEE}

\maketitle

\begin{abstract}
Contribution: The combination of ChatGPT with traditional learning resources is very effective in computer science education. High-performing students are the ones who are using ChatGPT the most. So, a new digital trench could be rising between these students and those with lower degree of fundamentals and worse prompting skills, who may not take advantage of all the ChatGPT possibilities.
Background: The irruption of GenAI such as ChatGPT has changed the educational landscape. Therefore, methodological guidelines and more empirical experiences in computer science education are needed to better understand these tools and know how to use them to their fullest potential.
Research Questions: This article addresses three questions. The first two explore the degree of use and perceived usefulness of ChatGPT among computer science students to learn database administration, where as the third one explore how the utilization of ChatGPT can impact academic performance.
Methodology: This contribution presents an exploratory and correlational study conducted with 37 students who used ChatGPT as a support tool to learn database administration. The student’s grades and a comprehensive questionnaire were employed as research instruments.
Findings: The obtained results indicate that traditional learning resources, such as teacher’s explanations and student’s reports, were widely used and correlated positively with student's grade. The usage and perceived utility of ChatGPT were moderate, but positive correlations between student’s grade and ChatGPT usage were found. Indeed, a significantly higher use of this tool was identified among the group of outstanding students. 
 
\end{abstract}



\section{Introduction}
\label{intro}

\IEEEPARstart{S}{tudents} are increasingly familiar with many Generative Artificial Intelligence (GenAI) tools and use them regularly. Some teachers tend to see it as a problem, because students have in their hands a tool that has the potential to create content that often they wouldn't do. Those teachers take for granted that their students will use it to cheat. The goal of experiments like the one presented in this article is to change this mindset and try to naturally apply the use of GenAI tools to enhance the learning experience. 

A remarkable quantity of sources, from renowned Universities to international institutions, are providing a set of recommendations for teachers and students to get along with this embracement process. Chronologically ordered, some of the most relevants are \cite{Galileo2023}, and a whitepaper \cite{Gimpel2023Unlocking}, taken as foundational for many of the following works. Followed by \cite{Gimpel2023Unlocking} or \cite{EDUCAUSE2023a}, applied by \cite{URJC2023} and \cite{UC3M2023} in Spanish ecosystem, or \cite{UnivCyprus2023}. Most especially, the UNESCO recommendations guide \cite{UNESCO2023}, a new EDUCAUSE guide more oriented to GenAI \cite{EDUCAUSE2023b}, the work from \cite{OpenUniv2023} or Russell Group recommendations \cite{Russell2023}. A recent compilation of studies in which ChatGPT and other GenAIs appear as new approaches for different educational purposes is shown in \cite{Koh2023}.

The next two subsections offer a summary of recommendations to use GenAI tools. The third subsection presents the related work that explore the usage of GenAI in computer science education, and more specifically, in database administration education. The last subsection introduces the present work and the posed Research Questions. 

\subsection{Recommendations for students}
Students use GenAI tools: in spring 2023, 27\% of students were using AI writing tools, and by the fall, that percentage had jumped to 49\%, despite some applied restrictions \cite{Tyton2023}. It is important to understand these tools, know how far they can go and integrate them into teaching and research effectively, responsibly, and ethically. That said, the most important recommendations for students can be listed as follows:

\begin{itemize}
\item Know and respect the regulations.
\item Focus on learning objectives: GenAI is a tool, not an end.
\item Use the GenAI: now is the time to learn how to use a GenAI tool correctly. For example: as a writing assistant, as a programming assistant, or as a teacher by running a conversation in Socratic mode. But being aware of the inherent risks in using GenAIs such as uploading personal information. Anyhow, the biggest risk is to use the answer that the GenAI returns at the first iteration, instead of offering a personal sieve and appealing to student critical spirit.
\end{itemize}  

Finally, whenever a GenAI had been used to perform a work, the recommendation is to declare it, like any other bibliography. Clues: which prompts have been used, and how the answers have been further processed.

\subsection{Recommendations for teachers}
As teachers put the GenAI to work at their service, using it with the due precautions, their teaching can be much more effective, reducing the time spent on a multitude of tedious tasks in order to focus on the important ones. All the recommendations split their application into the purely teaching and evaluation areas. 

In the area of teaching, the recommendations emphasize on providing rules and guides from the beginning of the course. While also keeping the focus on learning objectives, they promote using GenAI in a number of tasks (e.g., creating learning materials or designing a class, syllabus or topic). Moreover, certain GenAI can detect student profiles based on their performance, interests and abilities to generate exclusive, interactive and personalized content. Other goal should be a collaborative teachers-students learning on how to use GenAI. Involving the students in this process will increase the motivation of all players along the way. Other ways of animating learning could be to apply problem-based or challenge-based learning in scenarios determined by the GenAI, or to suggest prompts that end up challenging students to discover the limits of their knowledge, thus appealing the development of their critical thinking.

Regarding evaluation, a traditional approach, based on asking for a written work that students can do as homework, is a task that AI is going to create more and more perfectly. Thus, teachers must focus on what do they really want their students to learn: a concept, a skill, a specific competency? Exploring the student thinking process path could be increasingly important. Furthermore, GenAIs can help teachers to do their jobs much more effectively by aiding them in rubrics generation for an assessment; by providing quizzes in several formats; by contributing to the automatic correction of exercises, even those with elaborated answers; or by providing quick feedback paths to students on their work, automated to some extent, customized with the right GenAI training. It is therefore necessary to innovate in the assessments, both in the process, the format and the feedback that teachers can offer to the students.

\subsection{The Database Administration teaching with GenAIs}
GenAI is a tool that outbreaks as a disruptive influence on data management. \cite{DBLP:journals/pvldb/FernandezEFKT23} points out that it will impact in two ways for this area. First, several hard database problems such as entity resolution, schema matching, data discovery, and query synthesis, can be approached in a new way as the semantics of the data are out of the scope of automation. Thus, Large Language Models (LLMs) can aid to transform database tuples, schemes, and queries in real-world concepts. Second, these tools so prone to answer questions suppose a real connection with predictive models and information retrieval.

Exploring the use of AI in teaching for acquiring Database Administration (DBA) skills is not new. Twenty years ago \cite{Chau2003} reported, for instance, a case in which the students were impelled to create a proprietary search engine using two available tools in a semester course. After all these years we can find works like \cite{Cory2023}, which discusses the design process used in a graduate-level U.S. University advanced data analytics course including AI, focusing a part of its study on understanding AI impact on the accounting profession. They applied AI software (MindBridge) to analyze accounting data to evaluate risk, obtaining a 159\% increase in the mindset mean from no knowledge at the beginning of the course to average awareness at the end, which is a strong evidence of student learning close to learning objectives. Moreover, they show a remarkable improvement on critical thinking.

Focusing on the use of GenAI in software engineering, \cite{nguyenduc2023generative} provides a literature review in which up to 78 research questions were identified, showing the use of GenAI in a wide range of software development activities from design to education, but reckoning a lack of research to explore the steps to do. The collected experiences are related to the use of chatbots to assist the study, as well as the automatization of assessment to provide the students with immediate feedback.

There is an increasing number of papers discussing the possible use of GenAI in software engineering, although tests with students are still taking off. One important conclusion is that ChatGPT may enhance student programming self-efficacy and motivation, and the teachers must teach how to properly use it through prompt writing skills \cite{YILMAZ2023100147}. The results of \cite{educsci13090924} in a survey of 430 computer science Master Sc students suggest that many of them are familiar with ChatGPT but do not regularly use it for academic purposes, being skeptical of its positive impacts on learning unless guidelines and education on the tool is provided.

A teacher’s survey contextualized in nine courses at Univ. South-Eastern Norway can be seen in \cite{duc2023generative}. Three categories were explored: theory courses, programming courses (including DBA), and project-based courses. It was demonstrated that while GenAI tools can provide guidance on programming concepts up to an extent, students cannot fully develop their practical skills by using ChatGPT, especially those related to complex programming concepts or intricated problems like integrating libraries, specific system configurations, or large software systems. As some of the solutions provided by ChatGPT are outdated, it becomes clear that teacher assistant is highly desirable in combination with ChatGPT usage. 

Scarce and shy trials including GenAI tools in higher education DBA courses have appeared. University of Virginia has published a series of courses in which they include some examples of using AI in the classroom when it comes to learn statistics and Data Science \cite{UnivVirginia}. Despite it is a wonderful material, no results about their application are shared yet. 
There are some studies focused on how to teach SQL with ChatGPT. \cite{Balderas2022} at UNIR, a Spanish University, shows the use of a chatbot-based learning platform (based on IBM Watson) to iteratively assist students to perform SQL queries. Their results show that students who used the chatbot performed better on the final SQL exam (43\% pass) than those who did not (18\%). Interestingly, lecturers get illuminating metrics on student performance at the same time. \cite{Hong2023}, in a work in progress, are using ChatGPT for a similar purpose, but they have not yet reported concrete results. 

\subsection{Introducing this work}

\begin{figure}[ht]
	\centering
		\includegraphics[width=0.8\linewidth]{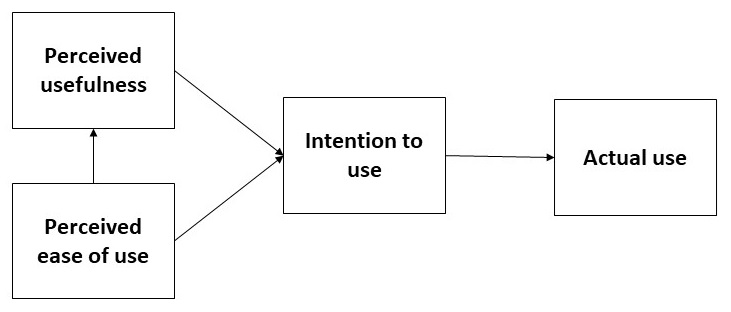}
	  \caption{Technology Acceptance Model, adapted from \cite{TAM2}}
        \label{TAM}
\end{figure}

In the light of the above previous results and attempts, we have conducted an exploratory and correlational study. This study uses the Technology Acceptance Model (TAM) as conceptual framework \cite{TAM1}. This model has been largely used in the last decades to analyze the acceptance of technologies \cite{TAM2} and serves to explore how the perceived usefulness and perceived ease of use affects to the actual use of a technology (see Figure \ref{TAM}).

The present paper addresses three Research Questions (RQ), the first two of which are inspired by the TAM model:  

\begin{itemize}
    \item RQ1. How widespread is the use of ChatGPT among computer science students to learn database administration?
    \item RQ2. How helpful is ChatGPT for computer science students to learn database administration?
    \item RQ3. Might the utilization of ChatGPT impact on the academic performance of computer science students who are learning database administration?
\end{itemize}

This study aims to push forward the experience on using GenAIs in teaching computer science, and more specifically database administration, following the previously described set of recommendations. The rest of the article is structured as follows. The materials and methods are described in Section 2. The results are presented and discussed, respectively, in Sections 3 and 4. Lastly, Section 5 presents the conclusions, limitations and future work drawn from this study.

\section{Research methodology}
\label{MM}

\subsection{Context}
This case study was carried out in a database administration (DBA) course. The course is part of the fifth semester of the bachelor’s degree in Technologies for the Information Society delivered by the Faculty of Computer Systems Engineering of the Universidad Politécnica de Madrid (UPM), located in Spain. This course is mandatory and accounts for 6 European Credit Transfer System credits (ECTS), equivalent to 150-180 hours of student work. The course deals with traditional database administration topics such as hardware configuration to storage databases, and database optimization, security, restoration or monitoring. During the course, students work in groups to perform a practical computer assignment. The case study presented in this paper is focused on the utilization and perceived utility of the resources employed by the participating students to complete the practical computer assignment as well as its corresponding individual exam.  

\subsection{Sample}
The sample compromises 37 students enrolled in the DBA course during the academic year 2023-24 who completed the practical assignment and the individual exam. They were 30 males (81\%) and 7 females (19\%), and the mean age was 21.54 with a standard deviation of 1.73. Regarding the attendance, 22 students (60\%) declared that they attended to all the practical classes, 12 declared that they attended to most of the classes (32\%), and 3 declared that they only attended some of the classes (8\%). All the students declared that had a free ChatGPT account (i.e., version 3.5) that they could use during the learning experience.

\subsection{Procedure and materials}
Once the theoretical part of the course was taught, students started the practical computer assignment in groups of three people. The students mainly used MySQL as database management system, and they operated it through console and MySQL Workbench. Secondarily, they also use MariaDB as database management system and operated it with phpMyAdmin. 

In the assignment, students mainly faced the following tasks: 1) configuration and creation of a MySQL database; 2) realization of an Extraction-Transformation-Load process from several MS Excel files; 3) optimization of queries using indexes; 4) creation of users and provision of permissions; 5) backup and recovery operations; 6) configuration and creation of a MariaDB database; 7) migration from MySQL to MariaDB database. The students performed the assignment in eight 2-hours practical lessons, also working autonomously out of class. To complete the assignment, the students submitted a group report in which the procedures employed to solve the above-mentioned tasks must be explained. 

The students were free to use any resource they consider to complete the practical assignment: notes and tutorials provided by the teacher, explanations of the teacher or colleagues, googling, traditional websites (e.g., stackoverflow, mysql official website, etc.), and GenAI systems (specifically, ChatGPT v.3.5). The students were encouraged to use all possible resources, and in the practical sessions the teacher exemplified the utilization of these resources solving some problems with their use.

Once the assignment was completed and the report was submitted, the students took the individual exam in a 1 hour and a half session. The exam compromised tasks similar to those outlined in the previously numbered list as 1, 2, 5, 6, and 7. To perform the individual exam the students were free to use any resource they consider, except for messaging applications that would allow them to communicate synchronously with other people. This included the assignment report, tutorials provided by the teacher, googling, traditional websites, and ChatGPT v.3.5.

Finally, once the students completed the assignment and took the individual exam, they fulfilled a questionnaire about the usage and the utility of the employed resources. The questionnaire was delivered online through the virtual learning environment of the course. Before submitting the questionnaire, students gave their informed consent to use the collected information for research purposes.

\subsection{Methods and instruments}
The first data gathered in this case study are the student's grade in the practical part of the course, which is the grade of the individual exam. This grade is scored from 0 to 10, and gives rise to the following scales: 0-4.9 (fail), 5-6.9 (pass), 7-10 (outstanding). The second data are the results of the questionnaire administered to collect students’ opinions about the utilized resources during the assignment and the exam realization. 

The questionnaire comprises three sections. The first section included questions about age, gender, and classroom attendance. The second section included statements about the usage and utility of some resources available for the students during the assignment, as well as a question about the purpose for which they used ChatGPT (if used). The third section was like the second one, but addressed the resources employed during the completion of the individual exam. In the statements about the usage and utility of the available resources the students should indicate a level using a Likert scale from one (nothing) to five (a lot). In the statements about the utility of these resources the students could also select the option ‘don't know/no answer’, provided when they did not use a certain resource. The questionnaire items are presented together with the results. 

\subsection{Data analysis}
The results of the student’s grades and the questionnaire were analyzed by using two descriptive statistics: the mean (M) and the standard deviation (SD). Moreover, inferential results were computed. To do so, the Kolmogorov-Smirnov test of normality was conducted to check the normality of the data, which resulted to be not normally distributed. Therefore, non-parametric statistical methods were used. 

First, the Kruskal-Wallis test was employed to compare and find out possible statistically significant differences on the usage and utility of the available resources among the students according to their grades. Moreover, the Eta Squared coefficient was employed to study the effect size of these differences. Regarding this coefficient, it must be considered that a value between 0.01 and 0.06 means a small effect size, a value between 0.06 and 0.14 means a medium effect size, and a value greater than 0.14 means a large effect size.

Second, the Spearman correlation test was employed to correlate the student’s grades with the student’s responses to the questionnaire items. Regarding the Spearman correlation coefficient (Rho), a positive value means a positive correlation and a negative value means a negative correlation. Moreover, a value lower than $|$0.1$|$ means there is no correlation, between $|$0.1$|$ and $|$0.3$|$ means a low correlation, $|$0.3$|$-$|$0.5$|$ a medium correlation, $|$0.5$|$-$|$0.7$|$ a high correlation, and greater than $|$0.7$|$ means a very high correlation.

\section{Results}
\label{results}

\subsection{Student’s grades}
\begin{figure}[ht]
	\centering
		\includegraphics[width=1\linewidth]{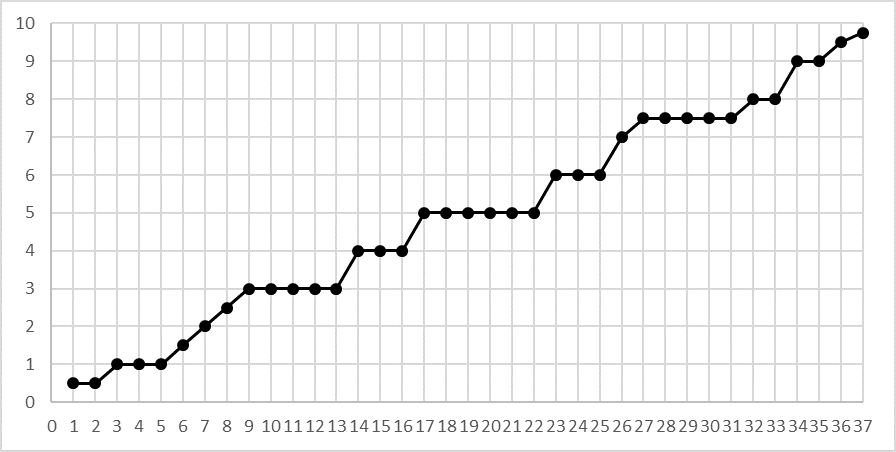}
	  \caption{Student's grades}\label{grades}
\end{figure}

Figure \ref{grades} depicts the students’ grades. The mean grade is 4.94 with a SD of 2.74. Of the 37 students, 16 failed the exam (grade lower than 5), 9 obtained a grade of pass (grade between 5 and 7) and 12 obtained a grade of outstanding (grade higher than 7). 

\subsection{Usage and perceived utility of resources}
\subsubsection{During the assignment realization}
The results obtained from the second section of the questionnaire show the usage and utility of some learning resources available to perform the assignment. The specific questions are: ‘To what extent have you used the following resources to perform the assignment?’ (usage), ‘How useful are the following resources to perform the assignment?’ (utility). Table \ref{tbl1} depicts these results, ordering the resources from highest to lowest usage. Moreover, the questionnaire also included a specific question about the purposes (that could be multi-selected from a closed list) for which students used ChatGPT during the realization of the assignment. Table \ref{tbl2} depicts these results, ordered by the number of occurrences. Note that 9 out of 37 students (around 25\%) did not use ChatGPT during the assignment realization and the remaining students (around 75\%) did use it.

\begin{table}[!t]
\caption{Usage and utility of resources during the assignment}\label{tbl1}
\centering
\begin{tabular}{|c|m{4cm}|m{1.3cm}|m{1.3cm}|}
\hline
& Item & Usage M(SD) & Utility M(SD) \\ \hline
1 & Assignment materials (i.e., teacher tutorials) & 4.28 (0.76) & 3.92 (1.20)\\ \hline
2 & Traditional internet resources (google, stackoverflow, etc.) & 4.08 (0.98) & 4.16 (1.01) \\ \hline
3 & Explanations of my group colleagues & 3.78 (1.18) & 3.92 (1.00) \\ \hline
4 & MySQL official documentation & 3.51 (1.19) & 3.47 (1.08)\\ \hline
5 & General teacher's explanations & 3.49 (1.30)	& 4.00 (0.97)\\ \hline
6 & Individualized teacher's explanations & 3.27 (1.19) & 4.23 (0.94)\\ \hline
7 & ChatGPT v3.5 & 3.03 (1.46) & 3.69 (1.17) \\ \hline
\end{tabular}
\end{table}

\begin{table}[!t]
\caption{Usage of ChatGPT during the assignment}\label{tbl2}
\centering
\begin{tabular}{|c|m{6cm}|m{1.2cm}|}
\hline
& Item & Occurrences \\ \hline
1 & To learn the syntax and examples of SQL statements & 18 \\ \hline
2 & To find a solution to a problem with a SQL statement & 15\\ \hline
3 & To look for alternatives or improvements to certain SQL statements & 13\\ \hline
4 & To identify SQL statements to solve a problem & 11\\ \hline
5 & I did not use ChatGPT & 9\\ \hline
6 & To be guided in the use of DBA tools & 5\\ \hline
7 & To identify Excel functionalities to solve a problem & 5\\ \hline
\end{tabular}
\end{table}

\subsubsection{During the exam realization}
The results obtained from the third section of the questionnaire show the usage and utility of some learning resources available to perform the individual exam. The specific questions are as follows: ‘To what extent have you used the following resources to perform the individual exam?’ (usage), ‘How useful are the following resources to perform the individual exam?’ (utility). Table \ref{tbl3} depicts the resources results ordered from highest to lowest usage. Moreover, the questionnaire also included a specific question about the purposes for which students used ChatGPT during the realization of the individual exam. Table \ref{tbl4} depicts these results, ordered by the number of occurrences. Note that 21 out of 37 students (around 55\%) did not use ChatGPT during the individual exam realization and the remaining students (around 45\%) did use it.

\begin{table}[!t]
\caption{Usage and utility of resources during the exam}\label{tbl3}
\centering
\begin{tabular}{|c|m{4cm}|m{1.3cm}|m{1.3cm}|}
\hline
& Item & Usage M(SD) & Utility M(SD) \\ \hline
1 & Assignment report & 4.49 (0.84) & 4.59 (0.76)\\ \hline
2 & Assignment materials (i.e., teacher tutorials) & 3.03 (1.36) & 3.46 (1.12)\\ \hline
3 & Traditional internet resources (google, stackoverflow, etc.) & 2.73 (1.56) & 3.53 (1.44)\\ \hline
4 & MySQL official documentation & 2.27 (1.33) & 2.96 (1.33)\\ \hline
5 & ChatGPTv3.5 & 2.14 (1.55) & 3.00 (1.54)\\ \hline
\end{tabular}
\end{table}

\begin{table}[!t]
\caption{Usage of ChatGPT during the exam}\label{tbl4}
\centering
\begin{tabular}{|c|m{6cm}|m{1.2cm}|}
\hline
& Item & Occurrences \\ \hline
1 & I did not use ChatGPT & 21\\ \hline
2 & To look for alternatives or improvements to certain SQL statements & 10\\ \hline
3 & To learn the syntax and examples of SQL statements & 8\\ \hline
4 & To find a solution to a problem with a SQL statement & 8\\ \hline
5 & To identify SQL statements to solve a problem & 4\\ \hline
6 & Others & 3\\ \hline
7 & To be guided in the use of DBA tools & 2\\ \hline
8 & To identify Excel functionalities to solve a problem & 1\\ \hline
\end{tabular}
\end{table}

\subsection{Usage and perceived utility of resources grouped by student’s grades}
The students could be grouped by grades according to the scale previously mentioned: fail, pass and outstanding. This leads to three groups: Group 1 (grade = fail, N=16), Group 2 (grade = pass, N = 9), and Group 3 (grade = outstanding, N = 12). The differences presented in the following Tables were analyzed by using the Kruskal Wallis test, remarking with a ‘*' symbol wherever a statistically significant difference appears.

\subsubsection{During the assignment realization}
Tables \ref{tbl5} and \ref{tbl6} depicts, respectively, the usage and utility of the available resources during the assignment. For each group the mean and the standard deviation (in parentheses) are displayed.

\begin{table}[!t]
\caption{Usage of resources during the assignment (grouped by grades)}\label{tbl5}
\centering
\begin{tabular}{|c|m{3.5cm}|m{1cm}|m{1cm}|m{1cm}|}
\hline
& Item & Group 1 & Group 2 & Group 3 \\ \hline
1 & Assignment materials (i.e., teacher tutorials) & 4.13 (0.70) & 4.67 (0.71) & 4.50 (0.80)\\ \hline
2 & Traditional internet resources (google, stackoverflow, etc.) & 4.25 (0.97) & 4.11 (1.27) & 3.83 (0.72)\\ \hline
3 & Explanations of my group colleagues	& 3.56 (1.17) & 4.33 (1.12) & 3.67 (1.15)\\ \hline
4 & MySQL official documentation * & 3.63 (0.93) & 4.22 (0.83) & 2.83 (1.40)\\ \hline
5 & General teacher's explanations * & 2.88 (1.41) & 4.67 (0.50) & 3.42 (0.90)\\ \hline
6 & Individualized teacher's explanations *	& 2.88 (1.17) & 4.78 (0.44) & 3.58 (0.79)\\ \hline
7 & ChatGPT v3.5 & 2.63 (1.17) & 3.33 (1.66) & 3.33 (1.61)\\ \hline
\end{tabular}
\end{table}

There are three differences identified as statistically significant at 0.05 level in Table \ref{tbl5}. In these items, the p-value and eta squared value are as follows: item 4: p-value = 0.03 / eta squared value = 0.13; item 5: p-value < 0.01, eta squared value = 0.27; item 6: p-value < 0.01 / eta squared value = 0.41.

\begin{table}[!t]
\caption{Utility of resources during the assignment (grouped by grades)}\label{tbl6}
\centering
\begin{tabular}{|c|m{3.5cm}|m{1cm}|m{1cm}|m{1cm}|}
\hline
& Item & Group 1 & Group 2 & Group 3 \\ \hline
1 & Assignment materials (i.e., teacher tutorials) * & 3.33 (1.35) & 4.78 (0.44) & 4.00 (0.95)\\ \hline
2 & Traditional internet resources (google, stackoverflow, etc.) & 3.94 (1.03) &	4.00 (1.32) & 4.58 (0.51)\\ \hline
3 & Explanations of my group colleagues	& 3.63 (1.11) & 4.44 (0.73)	& 3.91 (0.83)\\ \hline
4 & MySQL official documentation * & 3.07 (1.06) & 4.22 (0.67)	& 3.42 (1.08)\\ \hline
5 & General teacher's explanations * & 3.60 (1.05)	& 4.67 (0.71)	& 4.08 (0.67)\\ \hline
6 & Individualized teacher's explanations *	& 3.50 (0.91) & 5.00 (0.00) & 4.50 (0.67)\\ \hline
7 & ChatGPT v3.5 * & 3.09 (0.79) & 3.63 (1.30) & 4.40 (1.07)\\ \hline
\end{tabular}
\end{table}

Five differences identified as statistically significant at 0.05 level appear in Table \ref{tbl6}. In these items, the p-value and eta squared value are as follows: item1: p-value = 0.02 / eta squared value = 0.16; item 4: p-value = 0.03 / eta squared value = 0.43; item 5: p-value = 0.02, eta squared value = 0.17; item 6: p-value < 0.01 / eta squared value = 0.43; item 7: p-value = 0.01 / eta squared value = 0.17.

\begin{figure}[ht]
	\centering
		\includegraphics[width=1\linewidth]{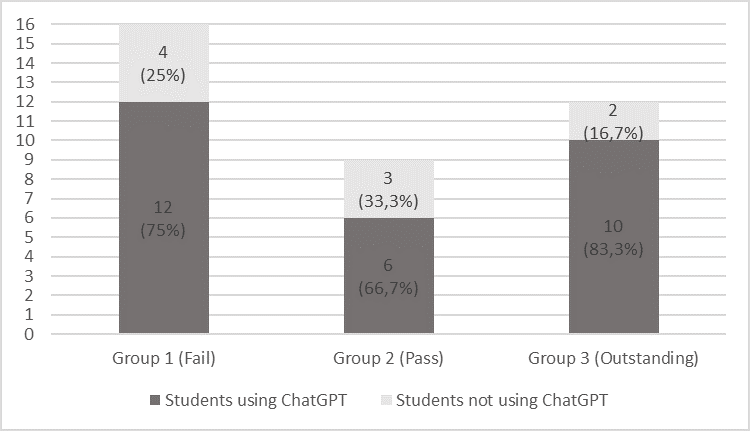}
	  \caption{Usage of ChatGPT during the assignment (results grouped by grades)}\label{usage_assignment}
\end{figure}

Lastly, the results regarding the item ‘I did not use ChatGPT (during the assignment)’ are analyzed by groups (Figure \ref{usage_assignment}). 

\subsubsection{During the exam realization}
Tables \ref{tbl7} and \ref{tbl8} depicts, respectively, the usage and utility of the available resources during the exam. For each group the mean and the standard deviation (in parentheses) are displayed.

\begin{table}[!t]
\caption{Usage of resources during the exam (grouped by grades)}\label{tbl7}
\centering
\begin{tabular}{|c|m{3.5cm}|m{1cm}|m{1cm}|m{1cm}|}
\hline
& Item & Group 1 & Group 2 & Group 3 \\ \hline
1 & Assignment report & 4.38 (0.69)	& 5.00 (0.00)	& 4.25 (1.13)\\ \hline
2 & Assignment materials (i.e., teacher tutorials) & 3.43 (1.41) & 2.66 (1.65) & 2.75 (0.86)\\ \hline
3 & Traditional internet resources (google, stackoverflow, etc.) & 3.00 (1.32) & 2.66 (2.00) & 2.41 (1.50)\\ \hline
4 & MySQL official documentation & 2.31 (1.04) & 2.88 (1.83) & 1.75 (1.05)\\ \hline
5 & ChatGPT v3.5 & 1.81 (1.23) & 1.66 (1.41) & 2.91 (1.78)\\ \hline
\end{tabular}
\end{table}

Although there appear to be some notable differences when applying Kruskal-Wallis test, none difference is identified as statistically significant at 0.05 level in Table \ref{tbl7}. Nevertheless, in the item 5 (the usage of ChatGPT) the resulting p-value is 0.14 and the eta squared value is 0.06.

\begin{table}[!t]
\caption{Utility of resources during the exam (grouped by grades)}\label{tbl8}
\centering
\begin{tabular}{|c|m{3.5cm}|m{1cm}|m{1cm}|m{1cm}|}
\hline
& Item & Group 1 & Group 2 & Group 3 \\ \hline
1 & Assignment report & 4.56 (0.60)	& 5.00 (0.00) & 4.33 (1.07)\\ \hline
2 & Assignment materials (i.e., teacher tutorials) & 3.75 (0.96)	& 3.42 (1.51)	& 3.08 (0.99)\\ \hline
3 & Traditional internet resources (google, stackoverflow, etc.) & 3.60 (1.20) & 3.25 (1.90) & 3.66 (1.41)\\ \hline
4 & MySQL official documentation & 2.88 (1.21) & 3.16 (1.72) & 3.00 (1.22)\\ \hline
5 & ChatGPT v3.5 * & 2.37 (0.99) & 2.20 (1.78) & 3.90 (1.37)\\ \hline
\end{tabular}
\end{table}

\begin{figure}[ht]
	\centering
		\includegraphics[width=1\linewidth]{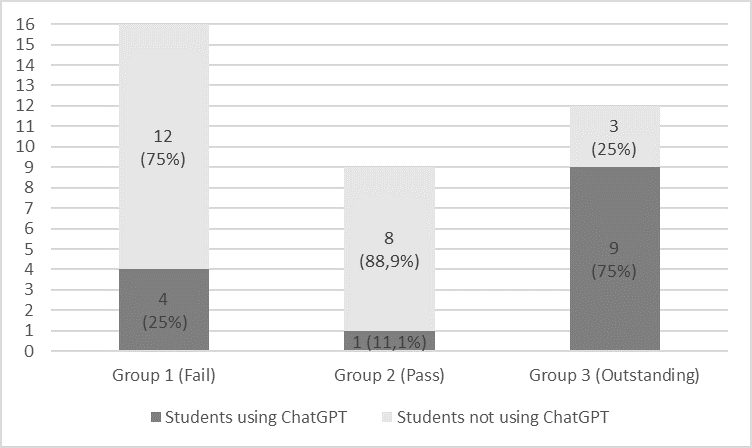}
	  \caption{Usage of ChatGPT during the exam (results grouped by grades)}\label{usage_exam}
\end{figure}

In Table \ref{tbl8}, there is a difference identified as statistically significant at 0.05 level in the item about the usage of ChatGPT, where the p-value is 0.04 and the eta squared value is 0.13. Figure \ref{usage_exam} shows the results regarding the item ‘I did not use ChatGPT (during the exam)’ analyzed by groups. 

\subsection{Relation between the student’s grades and the usage and utility of resources}
Correlations using the Spearman technique were computed in order to know how the student’s grades were related with the usage and perceived utility of the available resources during the assignment and the exam realization, as well as with the usage of ChatGPT. Table \ref{tbl9} depicts the assignment realization results, whereas Table \ref{tbl10} depicts the individual exam realization results. In both cases, the correlations are represented using the Spearman value (Rho) and the p-value. The correlations statistically significant at 0.05 level are marked with the ‘*’ symbol.

\begin{table}[!t]
\caption{Correlation between the student’s grade and the usage and utility of resources during the assignment realization}\label{tbl9}
\centering
\begin{tabular}{|c|m{4.5cm}|m{1cm}|m{1cm}|}
\hline
& Item & Usage & Utility \\ \hline
1 & Assignment materials (i.e., teacher tutorials) & 0.31 (0.06) & 0.26 (0.12)\\ \hline
2 & Traditional internet resources (google, stackoverflow, etc.) & -0.26 (0.11) & 0.23 (0.1)\\ \hline
3 & Explanations of my group colleagues & -0.06 (0.71) & 0.06 (0.71)\\ \hline
4 & MySQL official documentation & -0.14 (0.41) & 0.23 (0.17)\\ \hline
5 & General teacher's explanations & 0.23 (0.17) & 0.29 (0.08)\\ \hline
6 & Individualized teacher's explanations & 0.32 (0.05) * & 0.50 ($<$0.01) *\\ \hline
7 & ChatGPT v3.5 & 0.18 (0.28)	& 0.46 (0.01) *\\ \hline
\end{tabular}
\end{table}

\begin{table}[!t]
\caption{Correlation between the student’s grade and the usage and utility of resources during the exam realization}\label{tbl10}
\centering
\begin{tabular}{|c|m{4.5cm}|m{1cm}|m{1cm}|}
\hline
& Item & Usage & Utility \\ \hline
1 & Assignment report & 0.11 (0.51) & $\approx$ 0 (0.97) \\ \hline
2 & Assignment materials (i.e., teacher tutorials) & -0.33 (0.04) * & -0.27 (0.10)\\ \hline
3 & Traditional internet resources (google, stackoverflow, etc.) & -0.22 (0.19) & 0.01 (0.94)\\ \hline
4 & MySQL official documentation & -0.20 (0.22) & 0.12 (0.58)\\ \hline
5 & ChatGPT v3.5 & 0.22 (0.18) & 0.34 (0.10)\\ \hline
\end{tabular}
\end{table}

Moreover, there were computed the correlations between the student’s grade and the item ‘I did not use ChatGPT (during the assignment)’, with Rho = -0.13; p-value = 0.44, as well as between the student’s grade and the item ‘I did not use ChatGPT (during the exam)’, with Rho = -0.42; p-value = 0.01.

\section{Discussion}
\label{discussion}

\subsection{RQ1. How widespread is the use of ChatGPT among computer science students to learn database administration?}

In the light of the presented results, it can be said that GenAI are not so much extended among CS students to learn DBA if we compare it with other learning resources such as personal reports, ad-hoc tutorials made by the teacher, explanations of teachers and colleagues, or traditional websites (e.g., google, stackoverflow, etc.). Indeed, GenAI was the least used resource for both the assignment realization (see Table \ref{tbl1}) and the individual exam (see Table \ref{tbl3}). The resources most used by the students to perform the practical assignment and the exam were, respectively, the assignment materials (i.e., tutorials provided by the teacher) and the assignment group report (i.e., student’s resolution of the proposed tasks).

However, despite this, the usage of ChatGPT was moderately rated. Specifically, it was rated with a 3.03 (out of 5) and a 2.14 during the assignment and exam realization, respectively. Moreover, 75\% of the students used ChatGPT at some time to perform the practical assignment and 45\% used it to perform the individual exam. 

As the TAM model \cite{TAM1, TAM2} indicates, the use of a technology is influenced by its perceived ease of use and its perceived usefulness (which will be discussed in the next question). The ease of use of ChatGPT depends on having certain prompting skills, which the professor trained through the practice sessions by providing several examples to solve practical DBA problems using ChatGPT. Moreover, the ease of use  also depends on the user's knowledge of the topic he/she wishes to explore. In this last regard, the following observation is interesting.

The use of ChatGPT was not uniform among the students and it is relevant to observe the usage of this tool among students categorized on the basis of their grades. During the realization of the practical assignment (see Table \ref{tbl5}), a certain difference can be observed between the use made of ChatGPT by students who succeeded the exam (3.33 out of 5) and those who did not (2.63). Moreover, the proportion of students using ChatGPT was somewhat higher in the case of students who obtained a grade of outstanding (see Figure \ref{usage_assignment}). In particular, 83.3\% of the students in this group used ChatGPT, versus to 75\% and 66.7\% of the students who did it in the fail and pass groups. These differences are even greater if we examine the use of ChatGPT during the exam realization (see Table \ref{tbl7}). The difference continues to be in favor of students who obtained a grade of outstanding (2.91 out of 5) versus those who obtained a grade of pass (1.81) or fail (1.66). It is remarkable that in this case the difference had a medium effect size (eta squared value = 0.06). Moreover, by examining the proportion of students using ChatGPT among the different groups of students (see Figure \ref{usage_exam}), it can be observed that 75\% of the students in the outstanding group used ChatGPT, whereas the students of the fail and pass group used it much less (25\% and 11.1\%, respectively). These data suggest that the students who obtained lower grades did not achieve sufficient proficiency in ChatGPT for DBA problem solving during the assignment realization (which may be due to lack of DBA knowledge, lack of prompting skills, or both), so most of them decided not to use this tool during the exam.

Therefore, it can be concluded that although by now ChatGPT is not being used by CS students to learn DBA as much as other traditional learning resources, it is being moderately used. This conclusion is fully consistent with the work presented by \cite{educsci13090924}, where the usage of ChatGPT is still not so widespread by this student profile. Nevertheless, the usage of ChatGPT is not uniform and it is especially pronounced among students who obtain higher grades, which may indicate that: a) a proper use of ChatGPT help students to get better grades, b) high-performing students know how to make better use of ChatGPT, c) a combination of both, which sounds more reasonably.

\subsection{RQ2. How helpful is ChatGPT for computer science students to learn database administration?}
Regarding the usefulness of ChatGPT for learning about DBA as perceived by the students, the students did not perceive ChatGPT to be as helpful as the other learning resources as well. In fact, during the assignment realization (see Table \ref{tbl1}), this tool was perceived as less useful than the explanations of the teacher or colleagues, the tutorials provided by the teacher or traditional internet resources, while during the exam realization (see Table \ref{tbl3}), ChatGPT was perceived as less useful than the assignment report or traditional internet resources.

However, leaving aside the comparison of ChatGPT with other learning resources, the degree of usefulness of this tool is getting moderately high. In fact, the students rated with a score of 3.69 (out of 5) and 3.00 the degree of usefulness of ChatGPT for the realization of the assignment and the exam respectively. The students mainly used ChatGPT to learn the syntax and examples of SQL statements, to find a solution to a problem with a SQL statement, and to look for alternatives or improvements to certain SQL statements (see Tables \ref{tbl2} and \ref{tbl4}).

But once again, the perceived usefulness of ChatGPT was not uniform among the students and it is very interesting to compare the perceived utility of this tool among the students categorizing them on the basis of their grades. Regarding the assignment realization (see Table \ref{tbl6}), a significant statistically difference with a large size effect in the usefulness of ChatGPT was found. This difference is in favor of the students who obtained a grade of outstanding (rating 4.40), as opposed to those students who obtained a grade of pass or fail, who rated the usefulness of ChatGPT much lower (3.63 and 3.09, respectively). In the same vein, regarding the exam realization (see Table \ref{tbl8}), a significant statistically difference with a medium-to-large size effect was found. In this case, the students of the outstanding group rated ChatGPT utility with a 3.90, whereas the students of the pass and fail groups rated it with a 2.20 and 2.37 respectively. Furthermore, it is striking to note that for the students who obtained a grade of outstanding, ChatGPT was slightly more useful than other traditional resources. Meanwhile, the groups of students who obtained lower grades rated ChatGPT as the least useful resource, especially for taking the exam. This is consistent from the perspective of the TAM model \cite{TAM1, TAM2}, which states that ease of use (a factor that has been discussed in the previous RQ and, in the case of ChatGPT, is influenced by the user's prompting skills and knowledge of the topic to be explored) influences perceived usefulness.

Therefore, two main conclusions can be drawn from this discussion. First, in general terms ChatGPT is perceived by CS students as a moderately useful resource for learning about DBA, although not as useful as other traditional learning resources. This reinforces the results of the recent studies of \cite{duc2023generative} and \cite{YILMAZ2023100147}, showing that CS students demand further support from the teachers, rather than relying absolutely on the ChatGPT usage. It is also correlated with the \cite{educsci13090924} study in which the students declared that the ChatGPT usage could not be the only source to rely on. Nevertheless, our study shows an increase respect to these related works in the GenAI usage, as it is expected due to the increase of its knowledge and its widespread overall use. Second, students who obtain higher grades have a significantly different perception regarding the ChatGPT utility than students with lower scores and they perceive ChatGPT as a very useful tool, even more that other conventional resources. This may indicate that high-achieving students know how to take more advantage of ChatGPT (as commented before, due to their DBA knowledge, prompting skills or both) and therefore rate its usefulness more favorably, while lower-achieving students do not. The ChatGPT usage can also be related with an increment in students’ motivation, a result that appeared in \cite{YILMAZ2023100147} and that could be a feedback loop: higher use, higher motivation, deeper study, higher marks.

\subsection{RQ3. Might the utilization of ChatGPT impact on the academic performance of computer science students who are learning database administration?}
Analyzing Tables \ref{tbl5} to \ref{tbl8}, it can be said that there is a relation between the usage and perceived utility of ChatGPT and the student’s academic performance. To further explore this relation, the results correlating the students grades and the usage and perceived utility of ChatGPT will be taken into consideration. 

First, the assignment realization results (see Table \ref{tbl9}) show that there is a positive correlation between the student’s grade and the usage of the ChatGPT (Rho = 0.18). Similarly, it is also interesting to note how the item related to the non-use of ChatGPT correlates negatively with the student’s grade (Rho = -0.13). Nevertheless, the learner draws from many sources of knowledge and its academic performance is impacted by all of them. So, it is important to highlight other positive correlations found: individualized teacher’s explanations (Rho = 0.32), tutorials provided by the teacher (Rho = 0.31), and general teacher’s explanations (Rho = 0.23). Regarding the correlations of the student’s grade with the perceived utility of the resources, it is remarkable the high values obtained for individualized teacher’s explanations (Rho = 0.50) and ChatGPT (Rho = 0.46). All of this suggest that the teacher’s work (i.e., teacher’s explanations and materials) combined with the usage of ChatGPT during the assignment realization is highly effective as it leads to high performance.

Second, the exam realization results (see Table \ref{tbl10}) show that there is a positive correlation between the student’s grade and the usage of ChatGPT during the exam (Rho = 0.22). There is also a positive correlation between the student’s grade and the perceived usefulness of this tool (Rho = 0.34). In the same vein, it is also interesting to note the negative correlation between the student’s grade and the non-use of ChatGPT (Rho = -0.42), which is statistically significant at 0.01 level. Examining the results related to other learning resources, it can be observed another positive correlation between the student’s grade and the usage of the assignment report, which contained the student’s resolution of the tasks proposed in the assignment. In addition, negative correlations can be observed between the student’s grade and the rest of the available resources: teacher’s tutorials, traditional websites and MySQL official website. These resources were highly used and perceived as helpful during the assignment realization (see Table \ref{tbl1}), but the low use and perceived usefulness during the exam (see Table \ref{tbl3}) in conjunction with these negative correlations suggest that the individual exam, when time was limited, was not suitable for consulting these sources. However, it seems that the assignment report combined with a proper usage of ChatGPT was highly effective. These results, together with the aforementioned correlations between the ChatGPT use during the assignment and the student's grade, suggest that the students who used ChatGPT properly and sufficiently during the assignment realization and acquired the knowledge and skills to use ChatGPT as an aid to solve DBA problems had a powerful ally when they took the exam.

Therefore, it can be concluded that ChatGPT is a helpful tool to assist in the resolution of DBA problems and seems to positively impact on student performance. This is partially consistent with the work of \cite{Balderas2022}, in which improvements in academic performance were connected with the usage of GenAI. However, that work showed that lowest rated students improved their performance with the use of a chatbot, contrary to our conclusions. It could be a consequence of using a specific chatbot deployed for the subject, while ChatGPT is a general GenAI that seems to be more useful for higher skilled students. Nevertheless, the chicken or the egg causality dilemma arises here: are high-performing students better in a subject because they use ChatGPT? or do high-performing students know how to use ChatGPT better because they have a stronger knowledge of the subject? Interestingly, \cite{Balderas2022} declared that high-performing students suggested implementation of more complex items for their chatbot, which points toward the second direction. Here we can see how a new digital trench could be rising between students with higher skills and knowledge who can improve their results, compared with the ones with lower degree of fundamentals, who may not take advantage of all the ChatGPT possibilities.

Consequently, in order to reduce this trench and allow all students to get the most of GenAI, and in line with \cite{YILMAZ2023100147}, we deem that it is necessary to promote a higher-order critical thinking by prompting, and teaching how to prompt in a higher dimension is a must to be covered in the near future by teachers. It will provide the students with two different tools: improving their insight view of the subject, as well as enhancing their ability to extract information using GenAI as a powerful tool that they are going to use from now on in their professional environments.

\section{Conclusions}
\label{conclusions}

This paper has presented an empirical study with CS students about the usage of ChatGPT to learn DBA. The main findings are as follows: 

\begin{enumerate}
\item ChatGPT was moderately used, but not as much as other traditional learning resources. Likewise, ChatGPT was perceived as moderately useful, although not as much as other resources. 
\item Students who obtained higher grades used ChatGPT the most and found it to be a very useful resource, even more than some traditional learning resources.
\item Positive correlations were found between the student’s grade and the use and perceived usefulness of ChatGPT. Other learning resources correlated positively with student performance were: teacher’s explanations and tutorials, and student’s reports and notes.
\end{enumerate} 

These findings allow us to conclude that ChatGPT is a helpful educational tool for solving DBA problems that are strongly connected with high student performance. To do so, students should be trained on the usage of this tool and practice to master it. Nevertheless, ChatGPT should be combined with other traditional resources to learn DBA, being especially important the teacher's resources and explanations, as well as the reports or notes made by the own student during the practical part of the course. This is aligned with some recent studies, where the necessity of teacher’s assistance is recognized by the students. At the same time, other recent educational research works in this area show also that the best ranked students are also the ones who use ChatGPT more in their learning. These precedents are reinforced by our work. We deem that these conclusions, although strongly connected to CS education, are fully transferable to any engineering or science discipline in which the practical mastery of technologies and problem solving play a key role.

Finally, despite the positive outcomes, we recognized that this contribution is not free of limitations. First, the empirical study is not experimental, so no cause-and-effect relationships can be found. Therefore, it cannot be fully assured that using ChatGPT always leads to high academic performance neither that students who performed better do so exclusively thanks to ChatGPT. This limitation could be overcome by conducting an experimental study such as a randomized controlled trial where the ChatGPT usage was isolated, but this would entail many difficulties and it does not seem fair from a teaching perspective to allow ChatGPT usage only to certain students. Second, the sample size is quite small (i.e., 37), which undermines the generalization of the conclusions. Undoubtedly, more case studies like the one presented here (and, if possible, involving a larger sample) should be carried out in order to consolidate the conclusions. Moreover, the limited sample size prevents us to perform an analysis from a gender perspective, which could be also interesting.

\bibliographystyle{IEEEtran}
\bibliography{refs}

\newpage

\section{Biography Section}
 
\vspace{11pt}
\vspace{-33pt}
\begin{IEEEbiography}[{\includegraphics[width=1in,height=1.25in,clip,keepaspectratio]{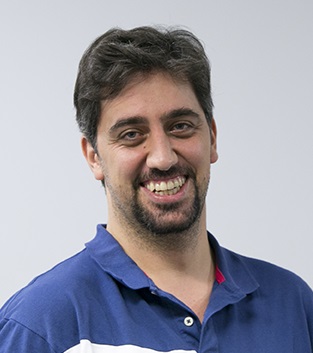}}]{Daniel López-Fernández} received the software engineer degree and the Ph.D. degree in software \& systems from the Universidad Politécnica de Madrid (UPM). He is currently Associate Professor with the Computer Science Department of the Computer Engineering School, UPM. His main research interests include the application of active learning methods, the development of innovative educational tools and the study of agile methodologies in software companies. He is currently the dean's delegate for Educational Innovation of the Computer Engineering School of the UPM and he is the director of the Educational Innovation Group of Educational Technologies and Active Learning Methods (GIETEMA). He has recently received the UPM Educational Innovation Award.

\end{IEEEbiography}

\vspace{11pt}

\begin{IEEEbiography}[{\includegraphics[width=1in,height=1.25in,clip,keepaspectratio]{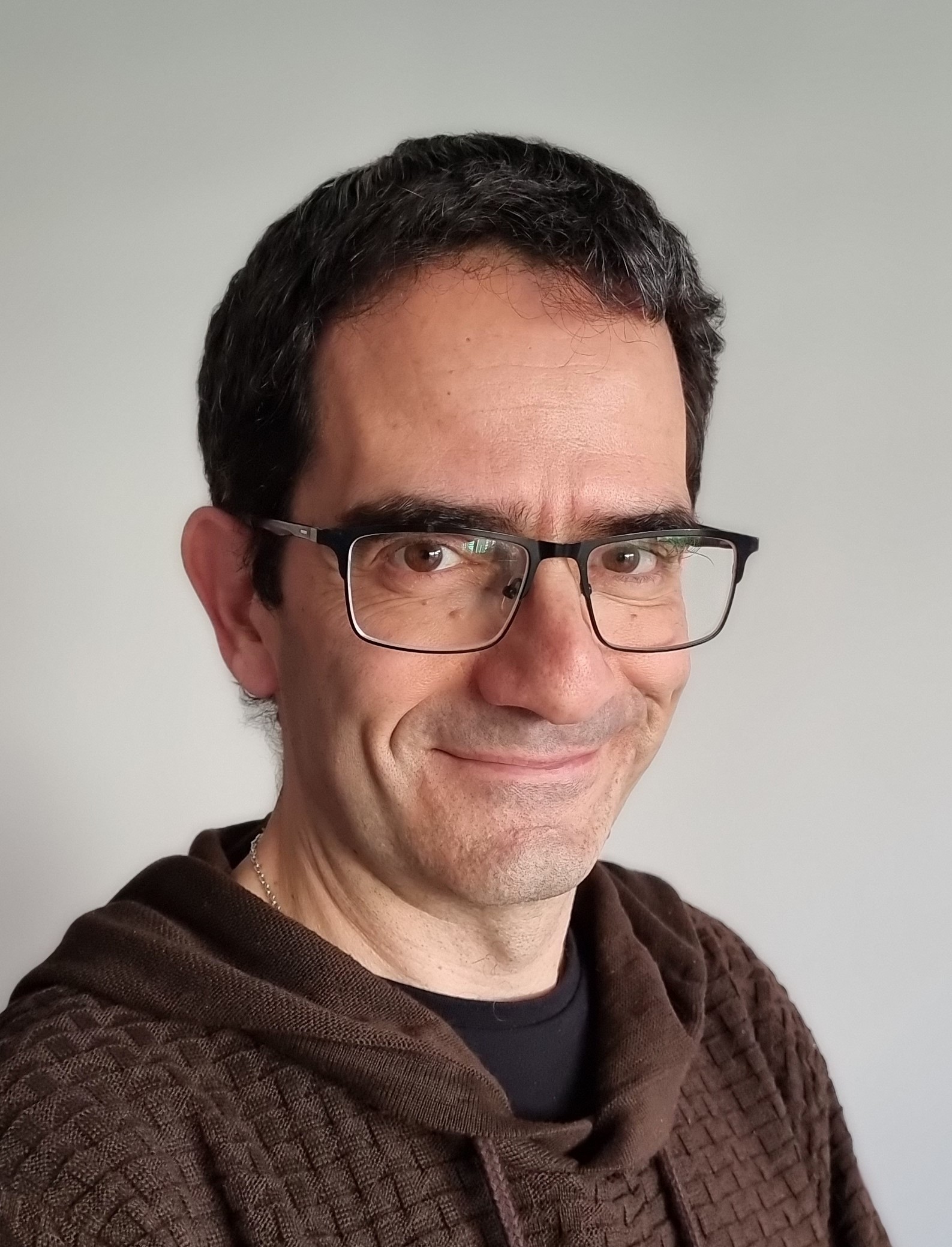}}]{Ricardo Vergaz Benito} received the electronics technology engineer degree and Physics M.Sc. and Ph.D. degree from Universidad de Valladolid. He is currently Associate Professor with Electronics Technology Department, Universidad Carlos III de Madrid, since 2001. His main research interests include assistive technologies, photonics devices and drivers for them. He has awarded several internal uc3m awards for his teaching, developed SPOCs and creating innovative content. He is also current Deputy Vice-rector of Infrastructures for uc3m digital, a unit devoted to innovative teaching and, recently, AI applied to it.

\end{IEEEbiography}

\vfill

\end{document}